\documentclass[useAMS, usenatbib, amssymb]{mn2e}
\usepackage{psfig}
\usepackage{graphicx}
\usepackage{epsf}
\usepackage{bm}
\usepackage{lscape}

\title [SDSS Luminosity colour relation for Thin Disc Stars]
{SDSS Absolute Magnitudes for Thin Disc Stars based on Trigonometric Parallaxes}

\author[Bilir et al.]
       {S. Bilir,${^1} \thanks{E-mail: sbilir@istanbul.edu.tr}$
        S. Karaali$^{2}$, S. Ak$^{1}$, K. B. Co\c skuno\u glu$^{1}$, E. Yaz$^{1}$
        A. Cabrera-Lavers$^{3,4}$
\\
  $^1$Istanbul University Science Faculty, Department of Astronomy and Space 
Sciences, 34119, University-Istanbul, Turkey\\
  $^2$Beykent University, Faculty of Science and Letters, Department of Mathematics  
and Computer, Beykent 34398, Istanbul, Turkey\\
  $^3$Instituto de Astrof\'{\i}sica de Canarias, E-38205 La Laguna, Tenerife, Spain\\
  $^4$GTC Project Office, E-38205 La Laguna, Tenerife, Spain\\}

\date{Accepted 2007 month day.
Received year month day; }

\pagerange{\pageref{firstpage}--\pageref{lastpage}} \pubyear{2007}

\begin{document}

\maketitle

\label{firstpage}

\begin{abstract}
We present a new luminosity-colour relation based on trigonometric parallaxes for thin disc main-sequence stars in SDSS photometry. We matched stars from the newly reduced Hipparcos catalogue with the ones taken from 2MASS All-Sky Catalogue of Point Sources, and applied a series of constraints, i.e. relative parallax errors ($\sigma_{\pi}/\pi\leq0.05$), metallicity ($-0.30\leq[M/H]\leq0.20$ dex), age ($0\leq t \leq 10$ Gyr) and  surface gravity ($\log g>4$), and obtained a sample of thin disc main-–sequence stars. Then, we used our previous transformation equations \citep{Biliretal08a} between SDSS and 2MASS photometries and calibrated the $M_{g}$ absolute magnitudes to the $(g-r)_{0}$ and $(r-i)_0$ colours. The transformation formulae between 2MASS and SDSS photometries along with the absolute magnitude calibration provide space densities for bright stars which saturate the SDSS magnitudes. 

\end{abstract}

\begin{keywords}
Galaxy: disc, Galaxy: solar neighbourhood, stars: distances
\end{keywords}

\section{Introduction}
Among several large sky surveys, two have been used most widely in recent years. The first, Sloan Digital Sky Survey \citep[SDSS,][]{York00}, is the largest photometric and spectroscopic survey in optical wavelengths. The second, Two Micron All Sky Survey \citep[2MASS,][]{Skrutskie06} has imaged the sky in infrared. SDSS obtains images almost simultaneously in five broad-bands ($u$, $g$, $r$, $i$ and $z$) centered at 3540, 4760, 6280, 7690 and 9250 \AA, respectively \citep[cf.][]{Fukugita96}. The photometric pipeline detects the objects, then matches the data from five filters and measures instrumental fluxes, positions and shape parameters. The shape parameters allow the classification of objects as "point source" or "extended". The limiting magnitudes of the passbands are 22, 22.2, 22.2, 21.3, and 20.5 for $u$, $g$, $r$, $i$, and $z$, respectively. The data are saturated at about 14 mag in $g$, $r$, and $i$, and about 12 mag in $u$ and $z$ \citep[cf.][]{Chonis08}.

2MASS provides the most complete database of near-infrared (NIR) galactic point sources. During the development of this survey, two highly automated 1.3m telescopes were used: one at Mt. Hopkins, Arizona to observe the Northern sky, and the other at Cerro Tololo Observatory, Chile to survey the Southern half. Observations cover 99.998 per cent \citep{Skrutskie06} of the sky with simultaneous detections in $J$ (1.25 $\mu$m), $H$ (1.65 $\mu$m) and $K_{s}$ (2.17 $\mu$m) bands up to limiting magnitudes of 15.8, 15.1 and 14.3, respectively. The passband profiles for $ugriz$ and $JHK_{s}$ photometric systems are given in Fig. 1 \citep{Biliretal08a}.

\begin{figure}
\begin{center}
\includegraphics[angle=0, width=92mm, height=80mm]{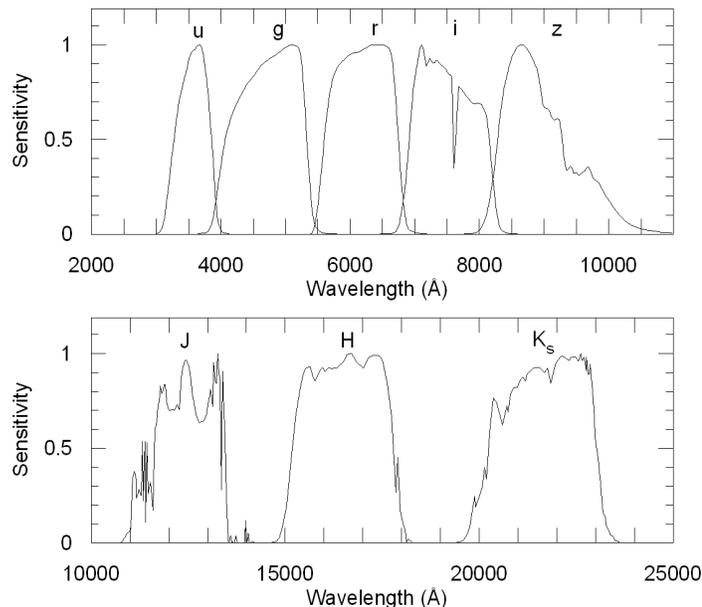}
\caption[] {Normalized passbands of SDSS filters (upper panel), and 2MASS filters (lower panel).}
\end{center}
\end{figure}

Saturation of the $ugriz$ data at bright magnitudes is a disadvantage for galactic surveys. Actually, space densities can not be evaluated for distances less than $\sim0.5$ kpc due to this saturation. The number of distance intervals without space densities is even larger for bright absolute magnitude intervals. On the other hand, 2MASS is a shallow survey, i.e. it contains relatively bright point sources, which can be used to fill up the gap of SDSS photometry for short distances. Our aim is to use the transformations between 2MASS and SDSS given in our previous paper \citep{Biliretal08a} and to calibrate the $M_{g}$ absolute magnitude using $(g-r)_{0}$ and $(r-i)_{0}$ colours based on Hipparcos trigonometric parallaxes. Thus, two sets of data will be used in estimating galactic model parameters: The first set is the $JHK_{s}$ data of relatively nearby stars in a given field of the Galaxy. The second one consists of SDSS data of stars occupying further distances in the same field. For the first set, 2MASS data needs to be transformed into SDSS data before applying procedures used in estimating absolute magnitude, whereas for the second set of data (the SDSS data) the procedures can be applied directly. 

In Section 2, the data and the determination of the sensitive sample are presented. The procedure and absolute magnitude calibration are given in Section 3 and the results are discussed in Section 4. Finally, a conclusion is given in Section 5.

\section{The data}

We matched stars from the newly reduced Hipparcos catalogue \citep{vanLeeuwen2007} with the ones from 2MASS All-Sky Catalog of Point Sources \citep{Cu03}, and applied a series of constraints in order to obtain a sample of thin disc main-sequence stars. To produce the sample, the first constraint we applied was to choose the 11644 stars from the newly reduced Hipparcos catalogue \citep{vanLeeuwen2007} with relative parallax errors $\sigma_{\pi}/\pi\leq0.05$. Then, we omitted the stars without 2MASS data. Afterwards, to eliminate reddening, 2MASS magnitudes were de–-reddened using the procedure given by \citet{Biliretal08a}, even though the program stars are relatively close and the near-infrared reddening correction is very small. The second restriction was to limit the absolute magnitude between $0<M_{J}<6$, which corresponds to the spectral type range A0--M0. The estimation of the absolute magnitude range can be seen from the 2MASS colour-magnitude diagram (Fig. 2). Finally, we adopted the procedure used in our previous paper \citep{Biliretal08b} to exclude evolved, thick disc and halo stars from the sample. The mentioned procedure requires the following limitations for metallicity, age and surface gravity taken from Padova isochrones \citep{Marigoetal08}: $-0.30\leq[M/H]\leq0.20$ dex, $0\leq t\leq10$ Gyr and $\log g>4$, respectively. Thus, to establish the luminosity--colour relation, we produced a thin disc sample of 4449 main–-sequence stars with accurate trigonometric parallaxes and 2MASS data. The typical near--infrared colour and absolute magnitude errors of our sample are $\pm0.04$ and $\pm0.14$ mag, respectively. \citet{LK73} stated that there is a systematic error in computed distances, which only depends on the $\sigma_{\pi}/\pi$ ratio. \citet{J01} showed that the error is negligible if $\sigma_{\pi}/\pi\leq0.10$, which is the case in this study as $\sigma_{\pi}/\pi\leq0.05$.

\begin{figure}
\center
\includegraphics[scale=0.35, angle=0]{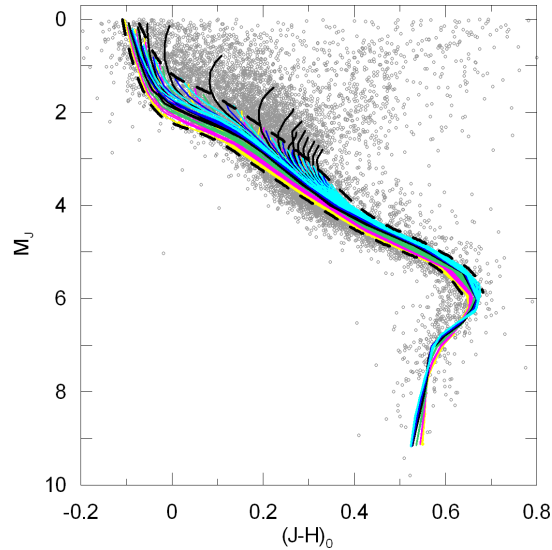}
\caption{$M_{J}/(J-H)_{0}$ colour-absolute magnitude diagram for the original 
sample. The upper and lower envelopes (the dashed lines) show the final sample, 
i.e. thin disc main--sequence stars. The thin curves correspond to Padova 
isochrones.}
\end{figure} 

\section{The procedure and absolute magnitude calibration}
We adopted the metallicity sensitive Eqs. (13), (18), and (19) from \citet{Biliretal08a} for the calibration of $M_{g}$ absolute magnitudes for our star sample. These equations are as follows:
\begin{eqnarray}
(g-J)_{0}=1.361(\pm0.016)(g-r)_{0}+1.724(\pm0.019)(r-i)_{0}\\
+0.521(\pm0.009),\nonumber 
\end{eqnarray}

\begin{eqnarray}
(g-r)_{0}=1.991(\pm0.040)(J-H)_{0}+1.348(\pm0.066)(H-K_{s})_{0}\\
–-0.247(\pm0.019),\nonumber  
\end{eqnarray}

\begin{eqnarray}
(r-i)_{0}=1.000(\pm0.036)(J-H)_{0}+1.004(\pm0.064)(H-K_{s})_{0}\\
-0.220(\pm0.017).\nonumber 
\end{eqnarray}

We transformed $(J-H)_{0}$ and $(H-K_{s})_{0}$ colours of the stars in our sample into $(g-r)_{0}$ and $(r-i)_{0}$ colours using Eqs. (2) and (3). Then, we used these colours and the $J_{0}$ magnitudes in Eq. (1) and obtained the $g_{0}$ magnitudes for the star sample. Finally, combining the $g_{0}$ magnitudes and Hipparcos parallaxes of stars we obtained accurate $M_{g}$ absolute magnitudes.

Now, we have two sets of SDSS data, i.e. $M_{g}$ absolute magnitudes and $(g-r)_{0}$ and $(r-i)_{0}$ colours for 4449 thin disc main--sequence stars. We can adopt a procedure similar to the one used in our recent works \citep{Bilir05,Biliretal08b} and calibrate the $M_{g}$ absolute magnitude to SDSS colours $(g-r)_{0}$ and $(r-i)_{0}$ as follows:

\begin{eqnarray}
M_{g}=a_{1}(g-r)^{2}_{0}+b_{1}(r-i)^{2}_{0}+c_{1}(g-r)_{0}(r-i)_{0}\\
+d_{1}(g-r)_{0}+e_{1}(r-i)_{0}+f_{1}.\nonumber
\end{eqnarray}

\begin{table*}
\setlength{\tabcolsep}{2pt} 
\center 
\scriptsize{
\caption{Coefficients and their standard errors for Eq. (4). $R^{2}$ and $s$ denotes the squared correlation coefficient and the standard deviation, respectively.}
\begin{tabular}{cccccccc}
\hline
 $a_{1}$ &  $b_{1}$  &  $c_{1}$ & $d_{1}$ & $e_{1}$ &  $f_{1}$ & $R^{2}$  &    $s$ \\
\hline
-0.719~($\pm$0.186) &  1.953~($\pm$0.681) & -0.474~($\pm$0.072) & 10.697~($\pm$0.196) & -9.350~($\pm$0.306) & 1.668~($\pm$0.027) & 0.99  & 0.19\\
\hline
\end{tabular}
} 
\end{table*}

We applied the covariance matrices of the solutions (1), (2) and (3) and obtained the individual estimation errors depending on the actual coefficients. These errors are given in the parentheses in Eqs. (1), (2) and (3) correspond to the dispersion of the observations. If we assume a total noise of approximately 0.1 associated with these calibrations, the complement error can be attributed to cosmic noise. 

The numerical values of the coefficients in Eq. (4) and their errors, the corresponding standard deviation and the squared correlation coefficient are given in Table 1. One can deduce that the error in absolute magnitude $M_{g}$ is of order 0.19 (the standard deviation). This is true when the colours in Eq. (4) are free of errors. However, this is not the case. The colours are associated with the errors given for the Eqs. (1), (2) and (3). Hence, one expects an additional error originating from the colours. We added the errors of the colours $(g-r)_{0}$ and $(r-i)_{0}$ to their observed values and reevaluated the absolute magnitudes by Eq. (4). For simplicity, we called absolute magnitudes evaluated using this procedure $M_{g}^{'}$. As expected, there is a mean offset of 0.12 mag from the Hipparcos absolute magnitudes. The standard deviation of $\Delta M_{g}^{'}=(M_{g})_{Hip}-M_{g}^{'}$ is 0.19. Hence, the total and maximum error for the absolute magnitude evaluated by Eq. (4) is $\sqrt{2}\times0.19=0.27$ mag.

\subsection{Testing the procedure}
We tested our procedure by comparing the $M_{g}$ absolute magnitudes calculated by means of Eq. (4) with the ones evaluated via combining $g_{0}$ apparent magnitudes and trigonometric parallaxes adopted from the newly reduced Hipparcos catalogue \citep{vanLeeuwen2007}. Fig. 3 shows that there is a one-–to--one correspondence between two sets of $M_{g}$ absolute magnitudes. Also, the mean and the standard deviation of differences inbetween original absolute magnitudes and calculated absolute magnitudes are rather small, i.e. $<\Delta M_{g}> \approx 0.00$ and $s=0.19$ mag.   

Most of the sample stars were obtained by applying a series of constraints to the newly reduced Hipparcos' catalogue. These stars lie within 2$\sigma$ (upper panel in Fig. 3). However, the $\Delta M_{g}=(M_{g})_{Hip}-(M_{g})_{c}$ residuals give the impression of a small, but systematic drift. That is, the calculated absolute magnitudes are a bit larger than the expected ones for faint absolute magnitudes. Keeping in mind the data sample were reduced from 11644 (original sample) to 4449, one can try to explain this drift with applied constraints. We should add that our aim in this work is to obtain absolute magnitudes for a pure thin disc population, so applying the constraints was needed. Hence, the results of the applied constraints are unavoidable. 

\begin{figure}
\center
\includegraphics[scale=0.35, angle=0]{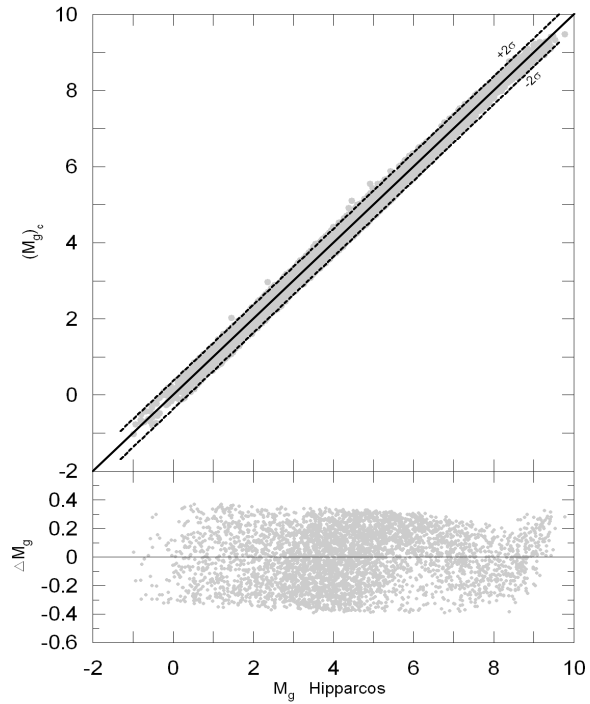}
\caption{($M_{g}$)$_{c}$ absolute magnitudes estimated using Eq. (4) versus $M_{g}$ absolute magnitudes 
calculated from newly reduced Hipparcos data (upper panel) and variation of the differences 
between two sets of absolute magnitudes (lower panel). All calibration stars in the 
figure are located within the prediction limit of $2\sigma$.}
\end{figure}

\subsection{Comparison of absolute magnitudes using different datasets}

We compared the $M_{g}$ absolute magnitudes determined in this work with the ones appearing in the literature. Comparison with the data of \citet{Covey07} and SEGUE data of SDSS DR6 by means of Allende Prieto et al.'s (2006) method could be carried out after some reductions and/or constraints, whereas a direct comparison could be made with \citet{KBT05}, \citet{Bilir05}, \citet{Juric08} and \citet{JH08}.

\subsubsection{Comparison with the data of Covey et al. (2007)}
\citet{Covey07} used the synthetic library of \citet{Pickles98} and evaluated the $M_{J}$ absolute magnitudes and, $(g-r)_{0}$ and $(r-i)_{0}$ colours for 18 main--sequence stars of different spectral types (Table 2). Using these values in the following equation (Eq. 5), modified from Eq. (1), we obtained the corresponding $M_{g}$ absolute magnitudes. Then, we compared these values with ($M_{g}$)$_{c}$ absolute magnitudes, which were calculated using Eq. (4). 
\begin{eqnarray}
M_{g}-M_{J}=1.361(\pm0.016)(g-r)_{0}+1.724(\pm0.019)(r-i)_{0}\\
+0.521(\pm0.009).\nonumber
\end{eqnarray}
There is an agreement between the two sets of absolute magnitudes (Table 2 and Fig. 4) confirming our new procedure for absolute magnitude determination. 
The small drift in the vertical direction in Fig. 4 for the faint absolute magnitudes originates from the application of two different procedures.  Although all the data, i.e. $M_{J}$, $(g-r)_{0}$ and $(r-i)_{0}$, were taken from \citet{Covey07}, the $(M_{g})_{c}$ absolute magnitudes (on the Y--axis) were evaluated only using $(g-r)_{0}$ and $(r-i)_{0}$ colours and Eq. (4), whereas $M_{g}$ absolute magnitudes (on the X--axis) were evaluated by substituting these colours and $M_{J}$ absolute magnitudes in Eq. (5).

\begin{table*}
\center 
\caption{Comparison of the absolute magnitudes calculated in our work with the absolute magnitudes of \citet{Covey07}. Data in columns (1-4) were taken from \citet{Covey07}. ($M_{g}$)$_{c}$ is calculated using Eq. (4) and the data in columns (3-4), $M_{g}$ is the absolute magnitude evaluated using Bilir et al.'s (2008a) procedure and the data in columns (2-4). $\Delta M_{g}$ is the difference between $M_{g}$ and ($M_{g}$)$_{c}$.}
\begin{tabular}{ccccccc}
\hline
Spectral Type &  $M_{J}$ & $(g-r)_{0}$ & $(r-i)_{0}$ &  $M_{g}$ & ($M_{g}$)$_{c}$ & $\Delta M_{g}$ \\
\hline
A0V & 0.430 & -0.250 & -0.180 & 0.300 & 0.674 & -0.374 \\
A2V & 1.170 & -0.230 & -0.170 & 1.085 & 0.797 & 0.288 \\
A3V & 1.250 & -0.160 & -0.150 & 1.295 & 1.373 & -0.078 \\
A5V & 1.380 & -0.100 & -0.110 & 1.575 & 1.638 & -0.063 \\
A7V & 1.730 & -0.020 & -0.080 & 2.086 & 2.214 & -0.128 \\
F0V & 2.430 & 0.100 & 0.010 & 3.104 & 2.637 & 0.467 \\
F2V & 2.630 & 0.190 & 0.030 & 3.461 & 3.393 & 0.068 \\
F5V & 2.620 & 0.260 & 0.030 & 3.547 & 4.118 & -0.571 \\
F6V & 2.900 & 0.280 & 0.080 & 3.940 & 3.861 & 0.079 \\
F8V & 2.980 & 0.360 & 0.100 & 4.163 & 4.493 & -0.330 \\
G0V & 3.180 & 0.380 & 0.140 & 4.460 & 4.333 & 0.127 \\
G5V & 3.540 & 0.490 & 0.160 & 5.004 & 5.254 & -0.250 \\
K2V & 4.500 & 0.780 & 0.240 & 6.496 & 7.354 & -0.858 \\
K3V & 4.940 & 0.850 & 0.320 & 7.170 & 7.320 & -0.150 \\
K4V & 5.210 & 1.000 & 0.380 & 7.747 & 8.195 & -0.448 \\
K5V & 5.450 & 1.180 & 0.400 & 8.267 & 9.638 & -1.371 \\
K7V & 5.770 & 1.340 & 0.540 & 9.046 & 9.888 & -0.842 \\
M0V & 5.720 & 1.310 & 0.640 & 9.127 & 8.866 & 0.261 \\
\hline
\end{tabular}  
\end{table*}

\begin{figure}
\center
\includegraphics[scale=0.34, angle=0]{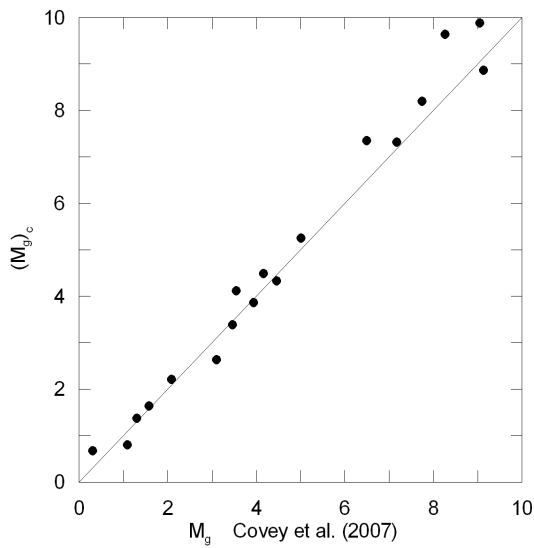}
\caption{($M_{g}$)$_{c}$ absolute magnitudes estimated using Eq. (4) versus $M_{g}$ absolute magnitudes calculated from Covey et al.'s data (2007).}
\end{figure}

\subsubsection{Comparison with SDSS DR6}
The second dataset we compared ours to is the Sloan Extension for Galactic Understanding and Exploration (SEGUE) data of SDSS Data Release 6 (DR6). Spectra for over 250000 stars in the galactic disc and spheroid for all common spectral types exist in SEGUE\footnote{http://cas.sdss.org/seguedr6/en/tools/search/sql.asp}. These spectra were processed with a pipeline called the ``Spectro Parameter Pipeline'' (spp) which computes standard stellar atmospheric parameters such as $[Fe/H]$, $\log g$ and $T_{eff}$ for each star by a variety of methods. We used the parameters evaluated by Allende Prieto et al.'s (2006) model-atmosphere analysis method and applied the following constraints in order to obtain a thin disc sample: $4\leq\log g\leq4.5$ and $-0.3\leq[M/H]\leq+0.2$ dex. The $M_{g}$ absolute magnitudes of 2289 stars, which satisfied our constraints, were evaluated by combining their $g_{0}$ apparent magnitudes and distances. The comparison of these absolute magnitudes with the ones determined using Eq. (4) , ($M_{g}$)$_{c}$, is shown in Fig. 5. There is an agreement between the two sets of absolute magnitudes. However, the dispersion is larger than the ones in Figs. 3 and 4. 
Also, there are clearly visible systematic effects shown in the distributions, in particular at $(M_{g})_{c}=6$. These effects probably originate from a series of constraints applied to the data, i.e. 1) we used the atmospheric parameters evaluated using Allende Prieto et al.'s (2006) method, 2) we selected stars with $4\leq \log g \leq4.5$, and 3) we limited the metallicities with $-0.3\leq[M/H]\leq+0.2$ dex in order to obtain a pure thin disc population. It seems that all these limitations caused the systematic effects mentioned above.

\begin{figure}
\center
\includegraphics[scale=0.34, angle=0]{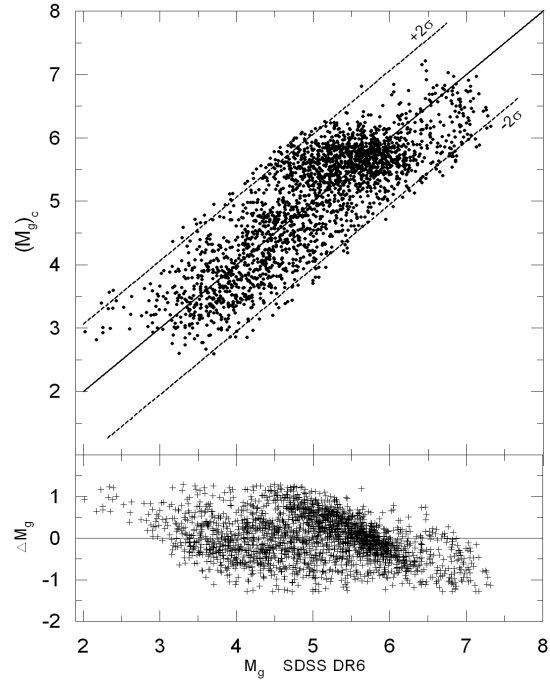}
\caption{($M_{g}$)$_{c}$ absolute magnitudes estimated using Eq. (4) versus $M_{g}$ absolute magnitudes calculated from SEGUE data of SDSS DR6 by means of Allende Prieto et al.'s (2006) method (upper panel) and variation of the differences between two sets of absolute magnitudes (lower panel). The dashed lines denote the $2\sigma$ prediction limits.}
\end{figure}
 
\subsubsection{Comparison with data appearing in the literature}
\citet{KBT05} used observations obtained in $u^{'}g^{'}r^{'}$ filters at Isaac Newton Telescope (INT) at La Palma in Spain. The filters were designed to reproduce the SDSS system. The data was complemented by \citet{L92} UBV standard star photometry and was used to calculate transformations between the INT SDSS $u^{'}g^{'}r^{'}$ filter–-detector combination and standard Johnson-–Cousins photometry. \citet{KBT05} presented transformation equations depending on two colours for the first time. The $M_{g}$ absolute magnitudes evaluated by these equations agree with the ($M_{g}$)$_{c}$ absolute magnitudes calculated using Eq. (4) in this work (Fig. 6). 
The  best fit in Fig. 6 is between the colour-–absolute magnitude diagrams obtained in our work and in the work of \citet{KBT05}. Actually, the open circle symbols used for the data of \citet{KBT05} and the locus of the grey circles representing the data in this work overlap.

\begin{figure}
\center
\includegraphics[scale=0.34, angle=0]{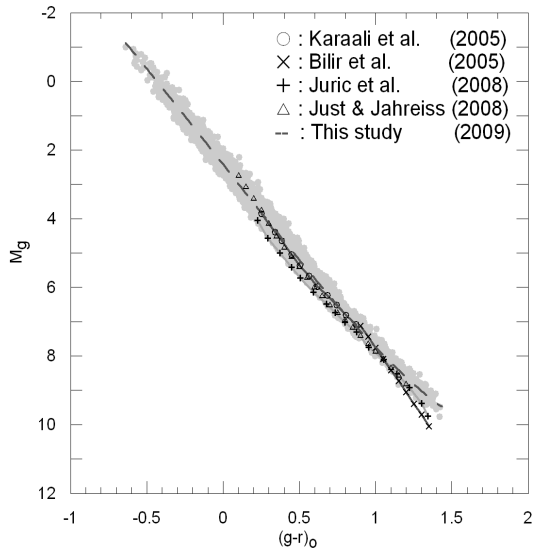}
\caption{$M_{g}/(g-–r)_{0}$ colour--magnitude diagram for different studies.}
\end{figure}

The transformations between SDSS and UBVRI photometries in our previous work \citep{Bilir05} provides $M_{g}$ absolute magnitudes for late-–type dwarfs. We compared the ($M_{g}$)$_{c}$ absolute magnitudes determined using Eq. (4) and the ones obtained using the procedure in \citet{Bilir05} to notice any possible deviation. Fig. 6 shows that the agreement is limited with colours $(g-r)_{0}<1.2$ mag.  
 
\citet{JH08} applied the transformation equations of \citet{Chonis08} to the 786 star sample in the solar neighbourhood ($r\leq25$ pc)  and obtained a mean $M_{g}/(g-–r)_{0}$ main sequence in 0.05 mag bins. The nearby stars represent the thin disc population completely. Hence, we expect an agreement between the $M_{g}$ absolute magnitudes evaluated in our work and in \citet{JH08}. Fig. 6 shows that this is the case. 

The only deviation is noticed between our $M_{g}$ absolute magnitudes and those of \citet{Juric08} (Fig. 6). However, their $M_{g}/(g-–r)_{0}$ colour--magnitude diagram is a combination of apparent colour-–magnitude diagrams and that may be the reason of the deviation mentioned (see Discussion).   

\subsection{Alternative procedure}
We applied  an alternative procedure to our sample of 4449 thin disc main–-sequence stars to evaluate the $M_{g}$ absolute magnitudes, explained as follows. First, we transformed their $(J-H)_{0}$ and $(H-K_{s})_{0}$ colours into $(g-r)_{0}$ and $(r-i)_{0}$ colours using Eqs. (2) and (3). Then, we used these colours and the corresponding $M_{J}$ absolute magnitudes, evaluated by the luminosity-–colour relations of Bilir et al. (2008b), in Eq. (5) and obtained the $M_{g}$ absolute magnitudes of the sample stars. We compared the $M_{g}$ absolute magnitudes evaluated by this procedure with the ones determined by the combining their apparent $g_{0}$ magnitudes and trigonometric parallaxes taken from the newly reduced Hipparcos catalogue \citep{vanLeeuwen2007}. Fig. 7 confirms the agreement between the two sets of $M_{g}$ absolute magnitudes. 
From the comparison of Fig. 3 and Fig. 7, one can deduce that the alternative procedure favours the procedure explained in Section 3. That is, the combination of $(g-r)_{0}$ and $(r-i)_{0}$ and $M_{J}$ absolute magnitude supplies more accurate $M_{g}$ absolute magnitudes by means of Eq. (5) than the combination of the same colours alone in Eq. (4). Actually, the mean of the offsets and the corresponding standard deviations for the data in Fig. 7 are $\Delta M_{g}=0.00$ and $s=0.03$, respectively, whereas they are $\Delta M_{g}\cong0.00$ and $s=0.19$ for the data in Fig. 3. The sharp limits of the residuals in the lower panel in Fig. 7 are due to concentration of the residuals with small standard deviation. We confess that, in order to obtain this figure, we omitted 36 sample stars with $\Delta M_{g}<-0.10$ mag which are not of 'AAA' quality flag.

\begin{figure}
\center
\includegraphics[scale=0.34, angle=0]{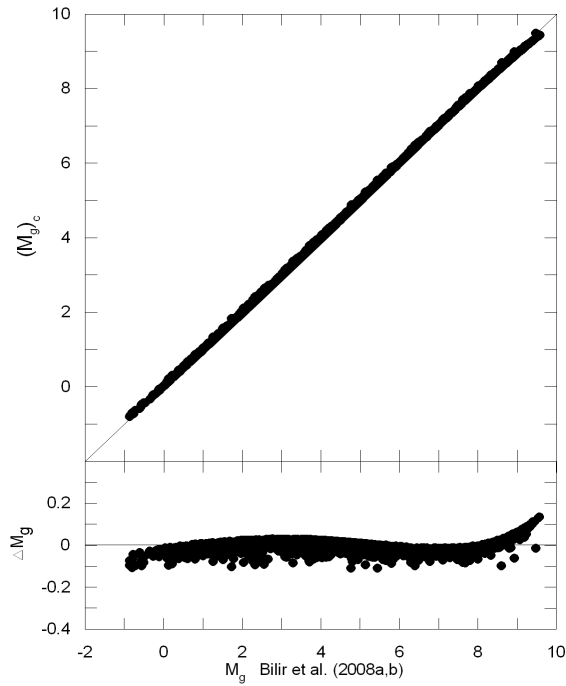}
\caption{$M_{g}$ absolute magnitudes estimated using Eq. (5) versus ($M_{g}$)$_{c}$ absolute magnitudes calculated from newly reduced Hipparcos data (upper panel) and variation of the differences between two sets of absolute magnitudes (lower panel).}
\end{figure}

\subsection{Binarism Effect}
A high fraction of stars are in fact binary systems and being a binary system makes stars appear brighter and redder than they normally are. Different fractional values (defined as $f$) can be found in the literature. As it was quoted in our previous work \citep[][and the references therein]{Biliretal08b}, the range of $f$ is $0.4\leq f < 1$. \citet{Kroupaetal91} found that for the extreme value, i.e. $f=1$, a single mass function provides the best representation of a single luminosity function. 

In our previous work \citep{Biliretal08b} we adopted a simple but reasonable procedure and revealed the binarism effect for $M_{V}$ and $M_{J}$ absolute magnitudes. The scatter in absolute magnitude as a function of binary fraction $f$ was found to be $\Delta M_{J}=0.266\times f+0.014$. In this work, the calibration of $M_{g}$ absolute magnitudes to $(g-r)_{0}$ and $(r-i)_{0}$ colours is carried out using the 2MASS data. Hence, we can adopt the same calibration for the scatter in $M_{g}$, i.e. 

\begin{equation}
\Delta M_{g}=0.266\times f+0.014.
\end{equation}

\section{Discussion}

We matched stars taken from the newly reduced Hipparcos catalogue \citep{vanLeeuwen2007} with the ones taken from 2MASS All-–Sky Catalogue of Point Sources \citep{Cu03}, and applied a series of constraints in order to obtain a sample of thin disc main-–sequence stars. These constraints reduced the original sample of 11644 stars with relative parallax errors $\sigma_{\pi}/\pi\leq0.05$ to 4449. Then, we used the 2MASS data in the transformation equations presented in our previous work \citep{Biliretal08a} and calibrated the $M_{g}$ absolute magnitudes to the SDSS colours, i.e. $(g-r)_{0}$ and $(r-i)_{0}$. 

The advantage of this procedure is that it uses a specific sample of nearby stars with small parallax errors. The constraints, i.e. $-0.30\leq[M/H]\leq 0.20$ dex, $0\leq t\leq10$ Gyr and $\log g>4$, exclude any contamination of evolved thin and thick disc stars and halo stars. Thus, we obtained a colour-–magnitude diagram which provides accurate $M_{g}$ absolute magnitudes for the thin disc main--sequence stars. The calibration can be used directly for faint stars, whereas for bright stars with saturated SDSS magnitudes one needs to transform the 2MASS data into SDSS data via our previous transformation formulae \citep{Biliretal08a}. 

We compared the $M_{g}/(g-–r)_{0}$ colour--magnitude diagram obtained in this study with six others appearing in the literature. The best fit is between our diagram and that of \citet{KBT05} which indicates that there are no systematic deviations between the in $u^{'}g^{'}r^{'}$ filters at INT and the original SDSS filters. 

Since \citet{JH08} used a sample of 786 stars in the solar neighbourhood, which have a much higher probability of consisting of pure thin disc main–-sequence stars, their results are precise. The precision of Just \& Jahreiss' (2008) results along with the agreement level between our data and theirs confirms our calibration. 

The $M_{g}$ absolute magnitudes evaluated from the synthetic data of \citet{Covey07} also agree with our $M_{g}$ absolute magnitudes. The same agreement holds for the $M_{g}$ absolute magnitudes of SEGUE data of SDSS DR6 by means of Allende Prieto et al.'s (2006) method, however, with larger dispersion. 

$M_{g}/(g–-r)_{0}$ colour--magnitude diagram of \citet{Juric08} deviates from our colour-–magnitude diagram and from those cited above. These authors compared the $M_{g}/(g–-r)_{0}$ colour--magnitude diagrams appearing in the literature and used two colour-–magnitude diagrams, one for bright stars and one for faint stars in their extensive work. We preferred the one for bright stars in this comparison due to our sample of stars being bright. However, we could not avoid the deviation mentioned. 

Finally, we should note that the $M_{g}/(g–-r)_{0}$ colour--magnitude diagram presented in this work agrees with the previous one \citep{Bilir05} for stars with $(g-r)_{0}<1.2$ mag, but there is a deviation for red stars which increases slightly with $(g-r)_{0}$ colour.

\section{Conclusion}
We calibrated the $M_{g}$ absolute magnitudes to the $(g-r)_{0}$ and $(r-i)_{0}$ colours by using the newly reduced Hipparcos catalogue and 2MASS data for thin--disc main--sequence stars with bright apparent magnitude which can be used in evaluating space densities for nearby stars. Thus, any possible degeneration due to extrapolation of density functions to zero distance should be avoided. This will lead to more accurate galactic model parameter estimation.         
               
\section{Acknowledgments}
We would like to thank the referee Floor van Leeuwen for his suggestions towards improving the paper. Also, we would like to thank Dr. Martin L\'opez-Corredoira for his valuable contributions. S. Karaali is grateful to the Beykent University for financial support.

\end{document}